\documentclass[twocolumn,aps,prb]{revtex4}

\usepackage{graphics}
\usepackage{epsfig}
\usepackage{color}
\usepackage{graphicx}
\usepackage{epsfig}
\usepackage{amsmath, amssymb}
\usepackage{verbatim}
\usepackage{bm}

\def\bra#1{\left\langle#1\right|}
\def\ket#1{\left|#1\right\rangle}

\begin{document}

\title{Topological nematic states and non-Abelian lattice dislocations} 

\author{Maissam Barkeshli and Xiao-Liang Qi}

\affiliation{Department of Physics, McCullough Building, Stanford
University, Stanford, CA 94305-4045}

\date{\today}

\begin{abstract}
An exciting new prospect in condensed matter physics is the possibility of realizing fractional quantum Hall (FQH) states in simple lattice models without a large external magnetic field. A fundamental question is whether qualitatively new states can be realized on the lattice as compared with ordinary fractional quantum Hall states. Here we propose new symmetry-enriched topological states, \emph{topological nematic states}, which are a dramatic consequence of the interplay between the lattice translation symmetry and topological properties of these fractional Chern insulators. When a partially filled flat band has a Chern number $N$, it can be mapped to an $N$-layer quantum Hall system. We find that lattice dislocations can act as wormholes connecting the different layers and effectively change the topology of the space. Lattice dislocations become defects with non-trivial quantum dimension, even when the FQH state being realized is by itself Abelian. Our proposal leads to the possibility of realizing the physics of topologically ordered states on high genus surfaces in the lab even though the sample has only the disk geometry.
\end{abstract}


\maketitle


\section{Introduction}

Some of the most important discoveries in condensed matter physics are the integer and fractional quantum Hall (IQH and FQH) states\cite{klitzing1980,tsui1982,laughlin1981,laughlin1983}, 
which provided the first examples of electron fractionalization in more than one dimension and paved the way for our current understanding of topological order \cite{zhang1992,wen04,nayak2008}.
Conventionally realized in two-dimensional electron gases with a strong perpendicular magnetic field, these states exhibit a bulk energy gap and topologically
protected chiral edge states. The topological order of the FQH states is characterized by topology-dependent ground state degeneracies\cite{wen1990} and fractionalized
quasi-particles \cite{laughlin1983}, 
while the quantized Hall conductance is determined by a topological invariant -- the Chern number -- which for a band insulator (IQH states)
can be determined by the momentum-space flux of the Berry's phase gauge field.\cite{thouless1982}.


Since Chern numbers are generic properties of a band structure, it is natural to expect that quantum Hall states can also be realized
in lattice systems without an applied magnetic field \cite{haldane1988,qi2005}, and indeed such ``quantum anomalous Hall'' (QAH) states
may be realizable experimentally \cite{liu2008B,yu2010b}. Recently, numerical evidence of fractional QAH (FQAH) states -- equivalent
to the $1/3$ Laughlin state and non-Abelian Pfaffian state --  has been found in interacting lattice models with (quasi-)flat bands
with Chern number $C = 1$ \cite{sun2011,regnault2011,wu2011,wang2011,neupert2011,tang2011,sheng2011,xiao2011}.  A flat band with $C = 1$ is similar to a Landau level,
where the kinetic energy is quenched and interaction effects are maximized. Recently, a systematic wave function approach was introduced
that demonstrates how to associate each FQH state in a Landau level to a counterpart in a generic $C = 1$ band \cite{qi2011}.

While known FQH states can be realized on the lattice (see also \cite{mcgreevy2011,parameswaran2011,vaezi2011,murthy2011}),
a fundamental question is whether new states can emerge as well.
Here we show that the answer is yes. There are two key reasons that the FQAH system is different: the possibility of a band with $C > 1$, and the interplay of
topological order with lattice symmetries \cite{wen2002,wen04,lu2011}. While each Landau level has $C = 1$, a single band in a lattice system
can in principle carry arbitrarily high Chern number. For example a model with $C=2$ quasi-flat bands has been recently proposed\cite{fwang2011}.
Here we show that there are new topologically ordered states -- \it topological nematic
states \rm -- in a partially filled band with $C > 1$. These states carry a nontrivial representation of the translation symmetry and
spontaneously break lattice rotation symmetry \cite{kivelson1998}. By using the Wannier function representation \cite{qi2011}, a band with Chern number
$C$ can be mapped to $C$ layers of Landau levels, but the $C$ layers are cyclically permuted under certain lattice translations.
This leads to a dramatic interplay of these states with the lattice translation symmetry: a pair of lattice dislocations connecting different layers
corresponds to a ``wormhole'' in the $C$-layer FQH system. The lattice dislocations effectively change the topology of the space,
giving rise to topological degeneracies. Surprisingly, the topological degeneracy associated with dislocations is
nontrivial even when the state itself, in the absence of dislocations, is an Abelian topological state \cite{barkeshli2010,bombin2010,kitaev2011,kitaev2006}.
Our result provides a new possibility of realizing exponentially large topological degeneracies without using
a ``genuine" non-Abelian FQH state, and effectively provides a way to experimentally observe the
topology-dependent ground state degeneracies of FQH states.

\section*{Wannier function description of Chern insulators}

We begin by reviewing the one-dimensional Wannier state description of Chern insulators\cite{coh2009,qi2011}.
For a band insulator with $N$ bands, the Hamiltonian is given by
\begin{eqnarray}
H=\int d^2{\bf k}c_{\bf k}^\dagger h({\bf k})c_{\bf k}\label{HQAH}
\end{eqnarray}
with $h({\bf k})$ a $N\times N$ matrix and the annihilation operator $c_{\bf k}$ an $n$-component vector. In
the current work we will focus on the systems with only one band occupied. Denoting the occupied band by
$\ket{{\bf k}}$, the Berry phase gauge field is $ a_i({\bf k}) \equiv -i\bra{\bf k}\partial/\partial k_i\ket{\bf k}$.
The one-dimensional Wannier states are defined by
\begin{eqnarray}
\ket{W(k_y,n)}=\int \frac{dk_x}{\sqrt{2\pi}}e^{-ik_xn}e^{i\varphi(k_x,k_y)}\ket{\bf k}\label{Wannierdef},
\end{eqnarray}
where \cite{qi2011} 
\footnote{Compared to the corresponding expression in Ref. \cite{qi2011}, we have added the first two terms in Eq. (\ref{wilson}) which only depends on $k_y$ and fixes the gauge choice between the Wannier states of different $k_y$.}:
\begin{eqnarray}
\varphi(k_x,k_y)&=&\frac{k_y}{2\pi}\int_0^{2\pi}a_y(0,p_y)dp_y-\int_0^{k_y}a_y(0,p_y)dp_y\nonumber\\
& &+\frac{k_x}{2\pi}\int_0^{2\pi} a_x(p_x,k_y)dp_x-\int_0^{k_x} a_x(p_x,k_y)dp_x\label{wilson}\nonumber\\
\end{eqnarray}
The phase choice $\varphi(k_x, k_y)$ is determined by the condition that the states be maximally localized
in the $x$-direction, which is determined by the eigenvalue equation $\hat{x}\ket{W(k_y,n)}=x_n\ket{W(k_y,n)}$,
where $\hat{x}$ is the $x$-direction position operator projected to the occupied band\cite{kivelson1982}. The formalism here can be generalized to a torus with finite $L_x,L_y$, in which case the
Wilson line integral is replaced by a discrete summation \cite{yu2011}.

\begin{figure}
\centerline{
\includegraphics[width=3.1in]{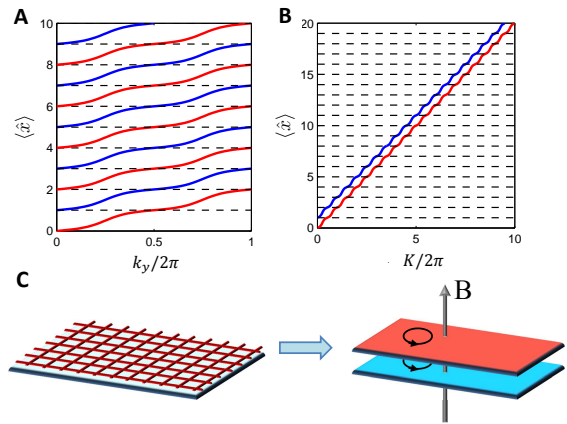}
}
\caption{
\label{fig:wannier}
 {\bf A}. Each state shifts over two lattice spacings: $\langle \hat{x} \rangle \rightarrow \langle \hat{x} \rangle + 2$, as $k_y \rightarrow k_y + 2\pi$.
Thus there are two families of states (red and blue lines). {\bf B}. The states can be mapped to two families of states, each parametrized by a single parameter $K_y$. {\bf C}. Illustration of the fact that the Chern number $2$ lattice system is mapped to a bilayer quantum Hall system, with the two layers corresponding to the two families of Wannier states shown in panel {\bf B}.
}
\end{figure}

The Wannier states form a complete basis of the occupied subspace, and each Wannier state is localized in the $x$
direction and a plane-wave in the $y$ direction. The essential property of the Wannier state is that its center of mass
position $x_n(k_y)=\bra{W(k_y,n)}\hat{x}\ket{W(k_y,n)}$, which is also the eigenvalue of the projected position operator
$\hat{x}$, is determined by the Wilson loop
\begin{eqnarray}
x_n(k_y)=n-\frac1{2\pi}\int_0^{2\pi}a_x(p_x,k_y)dp_x
\end{eqnarray}
Consequently, the Chern number \cite{thouless1982} of the band $C_1=\frac1{2\pi}\int dk_xdk_y\left(\partial_xa_y-\partial_ya_x\right)$
is equal to the winding number of $x_n(k_y)$ when $k_y$ goes from $0$ to $2\pi$:
\begin{eqnarray}
\int_0^{2\pi}\frac{\partial x_n(k_y)}{\partial k_y}dk_y=C_1
\end{eqnarray}
In other words the Wannier states satisfy the twisted boundary condition $\ket{W(k_y+2\pi,n)}=\ket{W(k_y,n+C_1)}$ and
correspondingly $x_n(k_y+2\pi)=x_n(k_y)+C_1$. Due to such a twisted boundary condition, the Wannier states
for $C_1=1$ can be labeled by one parameter $K=k_y+2\pi n$, and the Wannier state basis $\ket{W_K}$ is in
one-to-one correspondence with the Landau level wavefunctions in the Landau
gauge in the ordinary quantum Hall problem \cite{qi2011}. For $C_1=2$, we have $\ket{W(k_y+2\pi,n)}=\ket{W(k_y,n+2)}$,
so that the Wannier states on even and odd sites are two distinct families, as shown in Fig. \ref{fig:wannier} (a).
By adiabatic continuation of the momentum $k_y$ one can define
\begin{eqnarray}
\ket{W^1_{K=k_y+2\pi n}}&=\ket{W(k_y,2n-1)}\nonumber\\
\ket{W^2_{K=k_y+2\pi n}}&=\ket{W(k_y,2n)}\label{Wannier1ddef}
\end{eqnarray}
such that $\ket{W^{1,2}_K}$ are both continuous in the parameter $K$, and for the same $K$ they are related by a translation in the $x$ direction.
Compared with the $C_1=1$ case, one can see that each family of states $\ket{W^1_{K}}$ or $\ket{W^2_K}$ is topologically equivalent to the
Wannier states of a $C_1=1$ Chern insulator, which is in turn topologically equivalent to a Landau level quantum Hall problem. Therefore
the Wannier state representation when $C_1=2$ defines a map from the $C_1=2$ Chern insulator to a {\it bilayer} quantum Hall problem,
with the Wannier states on the even and odd sites mapped to two layers of Landau levels. In contrast to a typical bilayer quantum Hall
system, the two ``layers" are now related by a lattice translation. This is the key observation which leads to the topological nematic states
that we will propose below.

Multilayer FQH states can exhibit a rich variety of different topological states; recently,
a large classification of possibilities was developed and applied to two-component FQH states.\cite{BW1001a,BW1001b}
The simplest two-component generalizations of the Laughlin states are the $(mnl)$ states with the wavefunction \cite{halperin1983}
\begin{align}
\Phi\left(\left\{z_i\right\},\left\{w_i\right\}\right)&=\prod_{i<j}\left(z_i-z_j\right)^m\left(w_i-w_j\right)^n\prod_{i,j}\left(z_i-w_j\right)^l
\nonumber\\
&\cdot \exp\left[-\sum_i\left(\left|z_i\right|^2+\left|w_i\right|^2\right)/4l_B^2\right],
\end{align}
where $l_B$ is the magnetic length, $z_i = x_i + i y_i$ are the compex coordinates of the $i$th particle in one layer, and similarly $w_i$
are the complex coordinates for the other layer.  These states can be written down for the fractional Chern insulators
by simply passing to the occupation number basis, $\Psi(\{n_i^{I} \})$, where $n_i^{I}$ is the occupation number of
the $i$th orbital associated with the $I$th layer.
There are also intrinsically multilayer non-Abelian states\cite{BW1001a, BW1001b, AL0205,AR0149}.
In the current paper we will focus on the $(mml)$ states, where $m \neq l$ for incompressible states. 

\section{Interplay with lattice translation symmetry and dislocations}

A 2D lattice is invariant under two independent translation operations $T_x$ and $T_y$. 
Their action on the Wannier states defined in Eq. (\ref{Wannierdef}) and (\ref{Wannier1ddef}) is
\begin{eqnarray}
T_x |W_{K}^1\rangle &=& |W_K^2\rangle, \;\;\; T_x |W_K^2\rangle = |W_{K+2\pi}^1\rangle,
\nonumber \\
T_y |W_K^a \rangle &=& e^{iK} |W_K^a \rangle.\label{translation}
\end{eqnarray}
Thus $T_x$ exchanges the two sets of Wannier states but $T_y$ does not.



\begin{figure}
\centerline{
\includegraphics[width=3.5in]{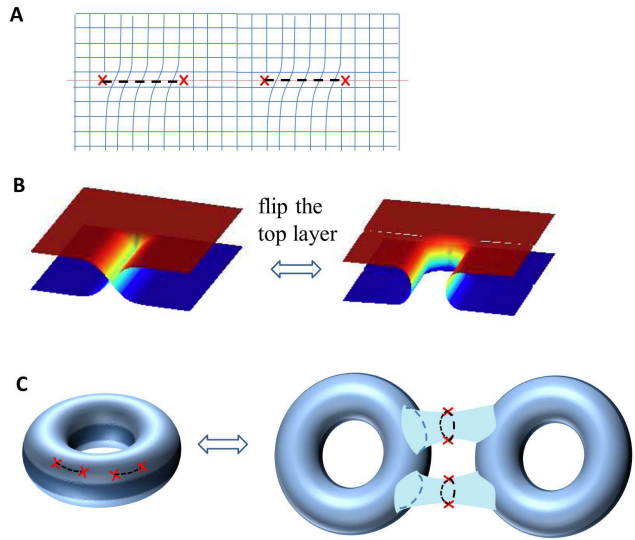}
}
\caption{
{\bf A}: Illustration of an $x$-dislocation. {\bf B}: (Upper pannel) Illustration that an $x$-dislocation leads to a branch cut around which the two effective layers are exchanged. (Lower pannel) A reflection of the top layer maps the branch cut between a pair of dislocations into a ``worm hole" connecting the two layers. {\bf C}: A torus with two pairs of $x$-dislocations is equivalent to two tori connected by two ``worm holes", which is a genus $3$ surface. This picture illustrates the fact that dislocations carry nontrivial topological degeneracy. 
\label{disloc} }
\end{figure}

Now consider the effect of dislocations\footnote{Several other topological effects induced by lattice dislocations have been studied in other topological states.\cite{juricic2011,ran2009}}; these are characterized by a Burgers vector $\bf b$,
which is defined as the shift of the atom position when a reference point is taken around a dislocation \cite{chaikin1995}. 
An $x$ dislocation with ${\bf b}=\hat{\bf x}$ is illustrated in
Fig. \ref{disloc} A. 
Far away from a dislocation, the lattice is locally identical to one without
a dislocation, so the dislocation is, as far as the structure of the lattice is concerned, a point defect.
Now consider a bilayer $(mmn)$ state realized on the lattice with a
dislocation. As is shown in Eq. (\ref{translation}), the two sets of Wannier states are related by translation in
$x$ direction. Thus when one goes around an $x$-dislocation, the two ``layers" consisting of Wannier states
$\ket{W_K^1}$ and $\ket{W_K^2}$ are exchanged. The map defined by the Wannier states, which
maps the $C_1=2$ Chern insulator to a bilayer FQH system, maps the Chern insulator on a lattice with a pair of
dislocations to a bilayer FQH state defined on a ``Riemann surface" with a pair of branch-cuts, as is illustrated in
Fig. \ref{disloc} B. This is the key observation which indicates that the $x$-dislocations in this system have nontrivial
topological properties. By comparison, the $y$-dislocations do not exchange the two layers and thus do not
correspond to a topology change in the effective bilayer description.

%
%

\section{Topological degeneracy of dislocations}

Although the $(mml)$ quantum Hall state considered is Abelian, the
$x$-dislocation carries a nontrivial topological degeneracy.\footnote{A related situation has been studied in two-component
Abelian quantum Hall states\cite{barkeshli2010} where $Z_2$ twist defects that have the effect of exchanging
QH layers were found to be non-abelian quasi-particles. Subsequently, $Z_2$ twist defects arising from
dislocations in the toric code model was studied in \cite{bombin2010}, where it was also found that they are
non-Abelian anyons. The current work is the first proposal where the $Z_2$ twist defect can be realized by a real lattice
dislocation in a realistic electronic system.} To understand this, start from the simplest case of $(mm0)$ state,
which is a direct product of two Laughlin states. For such a state, the Chern insulator on a torus is mapped to two
decoupled tori with a Laughlin $1/m$ state defined on each of them, with total ground state degeneracy of $m^2$. When a pair of
$x$-dislocations is introduced, the two tori are connected by the branch-cut. If we do a reflection
of the top layer according to the $x$ axis, the branch cut becomes a ``worm hole" between the two layers, as is illustrated in
Fig. \ref{disloc} B. Thus the two tori are connected, resulting in a genus $2$ surface. For two pairs of
dislocations, the two layers are connected by two worm holes and the whole system is topologically equivalent to a single Laughlin
$1/m$ state on a genus $3$ surface, as is shown in Fig. \ref{disloc} C. Thus the ground state degeneracy becomes $m^3$.
In general, when there are $2n$ $x$-dislocations on the lattice, the space is effectively a
genus $n+1$ surface and the ground state degeneracy for $n > 0$ is $m^{n+1}$. It follows that the average degree of freedom
carried by each dislocation--known as the quantum dimension--is $d=\sqrt{m}$. Thus we can see that the
$x$-dislocation carries a nontrivial topological degeneracy, in the same way as a non-Abelian topological quasiparticle.

The discussion above can be generalized: for the $(mml)$ state, $n > 0$ pairs of dislocations
on a torus leads to the topological degeneracy of $|m^2 - l^2| |m-l|^{n-1}$, so that the quantum dimension
of each dislocation is $d=\sqrt{|m-l|}$ (recall $m \neq l$ for incompressible FQH states). For $l\neq 0$, the system cannot be mapped to a Laughlin state on
high genus surface as the two layers are not decoupled \cite{barkeshli2010}. The topological degeneracy can be computed from
the bulk Chern-Simons effective theory \cite{barkeshli2010}.
In the following we provide an alternative understanding of the topological degeneracy using the edge states,
as it is more rigorous for $l \neq 0$, and it helps to provide a clearer understanding of the topological degeneracy.

\section{Topological degeneracy from edge state picture}\label{sec:edge}

\begin{figure}
\centerline{
\includegraphics[width=3.5in]{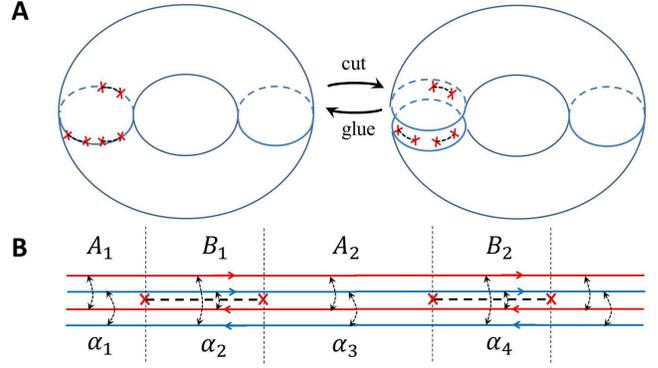}
}
\caption{The edge state understanding of the topological degeneracy. 
A. The dislocations are oriented along a single line, and then the system is cut along the line, yielding
gapless counterpropagating edge states along the line. The original FQH state is obtained from gluing the the
system back together by turning on appropriate inter-edge tunneling terms.
B. Depiction of the two branches (red and blue) of counter-propagating edge excitations. The arrows between the edge states
indicate the kinds of electron tunneling terms that are added. Away from the dislocations, in the
$A$ regions, the usual electron tunneling terms involving tunneling between the same layers, $\Psi_{eRI}^\dagger \Psi_{eLI} + H.c.$,
are added. In the regions including the branch cuts separating the dislocations, twisted tunneling terms are
added: $\Psi_{eR1}^\dagger \Psi_{eL2} + \Psi_{eR2}^\dagger \Psi_{eL1} + H.c.$. $\alpha_i$ indicate the mid-points of the $A$ or $B$ regions.
\label{fig1sup} }
\end{figure}

Here we will study in detail the topological degeneracy in the topological nematic states with dislocations
by using an edge state picture. By cutting the FQH state on a compact manifold along a line, one obtains a FQH state with
open boundaries and gapless counter-propagating chiral edge states on the boundary. The FQH state before the cut can be
obtained by ``gluing" the edge states back by introducing inter-edge electron tunneling (Fig. \ref{fig1sup}A). The topological degeneracy comes
from the fact that there are in general multiple degenerate minima when the electron tunneling becomes relevant \cite{wen1990edge}. In other
words, the topologically degenerate ground states of the bulk topological system correspond to degenerate ground states
due to spontaneous symmetry breaking in the edge theory.

The topological degeneracy of the $x$-dislocations is independent of their location, so to compute the topological
degeneracy we may orient the $n$ pairs of $x$-dislocations along a single line.
Next, we cut the system along this line to obtain a gapless CFT along the cut, and then we understand the resulting
ground state degeneracy by coupling the two sides of the cut in the appropriate way.

Below, we first review the chiral edge theory of Abelian FQH states and how to understand their torus degeneracy
from the point of view of the edge theory. Next, we show how this cut-and-glue procedure can be used to compute
the ground state degeneracy in the presence of dislocations for general two-component Abelian states.

The edge theory for an Abelian FQH state described by a generic $K$-matrix is given by the action \cite{wen1990edge,wen1992}
\begin{align}
\label{edgeAc}
S_{edge} = \frac{1}{4\pi} \int dx dt[ K_{IJ} \partial_t \phi_{LI} \partial_x \phi_{LJ} - V_{IJ} \partial_x \phi_{LI} \partial_x \phi_{LJ} ],
\end{align}
where $\phi_{LI}$ denotes left-moving chiral bosons for $I = 1, \cdots, \text{dim } K$. Here and below the repeated indices $I,J$ are summed. The field $\phi_{LI}$ is a compact
boson field with radius $R = 1$:
\begin{align}
\phi_{LI} \sim \phi_{LI} + 2\pi.
\end{align}
Quantizing the theory in momentum space yields \cite{wen04}
\begin{align}
[\partial_x \phi_{LI} (x), \phi_{LJ}(y)] = 2\pi K^{-1}_{IJ} \delta (x - y).
\end{align}
Integrating the above equation gives:
\begin{align}
[\phi_{LI}(x), \phi_{LJ}(y)] = \pi K^{-1}_{IJ} sgn(x-y)
\end{align}
The electric charge density associated with $\phi_{LI}$ is given by
\begin{align}
\rho_{LI} = \frac{1}{2\pi} \partial_x \phi_{LI},
\end{align}
and the $I$th electron operator is described by the vertex operator
\begin{align}
\Psi_{eLI} = e^{i K_{IJ} \phi_{LJ}} .
\end{align}
Note that normal ordering will be left implicit (\it ie \rm $e^{i K_{IJ} \phi_{LJ}} \equiv : e^{i K_{IJ} \phi_{LJ}} :$).
If we consider the FQH state on a cylinder, we will have a left-moving chiral theory on one edge, and a right-moving
chiral theory on the other edge. For the right-moving theory, the edge action is
\begin{align}
\label{RedgeAc}
S_{edge} = \frac{1}{4\pi} \int dx dt [-K_{IJ} \partial_t \phi_{RI} \partial_x \phi_{RJ} - V_{IJ} \partial_x \phi_{RI} \partial_x \phi_{RJ} ] ,
\end{align}
the charge is
\begin{align}
\rho_{RI} = \frac{1}{2\pi} \partial_x \phi_{RI},
\end{align}
and the electron operator is
\begin{align}
\Psi_{eRI} = e^{-i K_{IJ} \phi_{RJ}} .
\end{align}
Now if we bring the edges close together, the electrons can tunnel from one edge
to the other. The electron tunneling operators are:
\begin{align}
\frac{1}{2}\sum_I t_I [ \Psi_{eLI}^\dagger \Psi_{eRI} + \Psi_{eRI}^\dagger \Psi_{eLI} ] = \sum_I t_I \cos (K_{IJ} \phi_J),
\end{align}
where
\begin{align}
\phi_J = \phi_{LJ} + \phi_{RJ}
\end{align}
is a non-chiral boson. Note that
\begin{align}
[\phi_I (x), \phi_J(y)] = 0.
\end{align}
The tunneling terms gap out the edge and lead to a set of degenerate minima. Assuming that $t_I < 0$, the minima occur when
\begin{align}
K_{IJ} \phi_J = 2\pi p_I,
\end{align}
where $p_I$ is an integer. That is, the minima occur when $\phi_I = 2\pi K^{-1}_{IJ} p_J$.
Note that since $\phi_I \sim \phi_I + 2\pi$, the degenerate states can be labelled by
an integer vector $\v{p}$, which denotes the eigenvalues of of $e^{i \phi_I}$:
\begin{align}
\langle \v{p} | e^{i \phi_I} |\v{p} \rangle = e^{2\pi K^{-1}_{IJ} p_J},
\end{align}
for integers $p_I$. Two different vectors $\v{p}$ and $\v{p}'$ are equivalent if taking
$\phi_I \rightarrow \phi_I + 2\pi n_I $, for integers $n_I$ takes $\v{p} \rightarrow \v{p} + K \v{p} =  \v{p}'$.
Thus the ground state degeneracy on a torus is given by the number of inequivalent integer vectors
$\v{p}$, which is determined by $|\text{Det }K|$. \cite{wen1989}

For what follows, we specialize to the case where $K$ is a $2\times 2$ matrix,
$K = \left( \begin{matrix} m & l \\ l & m \end{matrix} \right)$. On a torus, there are $|\text{Det } K| = |(m-l)(m+l)|$
different ground states. From the edge theory, these can be understood
as eigenstates of $e^{i (\phi_1 \pm \phi_2)}$, with eigenvalue $e^{2 \pi i p_{\pm}/(m\pm l)}$, where $p_{\pm}$ are integers.
Therefore, the eigenstates can be labelled by $(p_+, p_-)$, where
\begin{align}
\langle (p_+, p_-) | e^{i (\phi_1 \pm \phi_2)} | (p_+, p_-) \rangle = e^{2 \pi i p_{\pm}/(m\pm l)}
\end{align}

Now suppose we have $n$ pairs of dislocations, such that going around each dislocation exchanges the
two layers, as discussed in the text. Each pair of dislocations is separated by a branch cut.
Let us align all the dislocations, and denote the regions without a branch cut as $A_i$, and the
regions with a branch cut as $B_i$ (Fig. \ref{fig1sup}A). Now imagine cutting the system along this line, introducing counterpropagating
chiral edge states. The gapped system with the dislocations can be understood by introducing different
electron tunneling terms in the $A$ and $B$ regions (Fig. \ref{fig1sup}B):
\begin{align}
\label{tunnL}
\delta \mathcal{L}_{tunn} =
\frac{g}{2} \left\{ \begin{array}{ccc}
\Psi_{eL1}^\dagger \Psi_{eR1} + \Psi_{eL2}^\dagger \Psi_{eR2}  + H.c & \text{ if } & x \in A_i \\
\Psi_{eL1}^\dagger \Psi_{eR2} + \Psi_{eL2}^\dagger \Psi_{eR1} + H.c  & \text{ if } & x \in B_i \\
\end{array} \right.
\end{align}
Introducing the variables
\begin{align}
\tilde{\phi}_1 &= \phi_{L1} + \phi_{R2}\nonumber\\
\tilde{\phi}_2&= \phi_{L2} + \phi_{R1}\label{phitilde}
\end{align}
we rewrite (\ref{tunnL}) as
\begin{align}
\delta \mathcal{L}_{tunn} =
g \left\{ \begin{array}{ccc}
\sum_I \cos(K_{IJ} \phi_J) & \text{ if } & x \in A_i \\
\sum_I \cos( K_{IJ} \tilde{\phi}_J) & \text{ if } & x \in B_i \\
\end{array} \right.
\end{align}

We can write the ground state approximately as a tensor product over degrees of freedom in the different regions:
\begin{align}
|\psi \rangle = \otimes_{i=1}^n |a_i \rangle |b_i \rangle,
\end{align}
where $|a_i\rangle$ is a state in the Hilbert space associated with $A_i$, and similarly
for $B_i$. The set of ground states is a subspace of the $| \text{Det } K|^{2n}$ states associated with
the minima of $\delta \mathcal{L}_{tunn}$ for each region.


If the interaction between different regions are ignored, one would have obtained $\left|m^2-l^2\right|$ degenerate states for each region labeled by eigenstates of $(\phi_1(x)\pm \phi_2(x))$ or $(\tilde{\phi}_1(x) \pm \tilde{\phi}_2(x))$ in each region. However, we observe that phases in $A$ and $B$ regions cannot be simultaneously diagonalized. Instead,
\begin{align}
\left[\phi_I(x),\tilde{\phi}_J(x')\right]=\frac{\pi}{m-l} \left(\begin{array}{cc}1&-1\\-1&1\end{array}\right)_{IJ}{\rm sgn}(x-x')
\end{align}
Therefore
\begin{align}
e^{i\phi_1(x)}e^{i(\tilde{\phi}_1(x')-\tilde{\phi}_2(x'))}e^{-i\phi_1(x)}&=e^{i\frac{2\pi}{m-l}{\rm sgn(x-x')}}e^{i(\tilde{\phi}_1(x')-\tilde{\phi}_2(x'))}\nonumber\\
\left[e^{i\phi_I(x)},e^{i(\tilde{\phi}_1(x')+\tilde{\phi}_2(x'))}\right]&=0
\end{align}
From these commutation relations we see that $\tilde{\phi}_1 + \tilde{\phi}_2 $ can have common eigenstates with $\phi_1$ and $\phi_2$ in the
$A$ regions, but the eigenstates of $\phi_1$ and $\phi_2$ are necessarily a superposition of the $m-l$ eigenstates of $\tilde{\phi}_1-\tilde{\phi}_2$.
From this consideration, we see that we can pick a basis where $|\text{Det } K|^n$ possible
states are associated with the $A$ regions, and the $B$ regions contribute $(m+l)^n$ possible states, for a total
\begin{align}
N_0=| \text{Det } K|^n (m + l)^n=(m^2-l^2)^n (m + l)^n
\end{align}
possible states.

However, there are also $2n-1$ global constraints that must be satisfied. 
%
The charge at each dislocation should be a definite quantity, which, without loss
of generality, we set to 0. This is because states at the dislocation which carry different charge
cannot be topologically degenerate, since charge is a local observable. 
These constraints are expressed as
\begin{align}
Q_i=\frac{1}{2\pi} \int_{\alpha_i}^{\alpha_{i+1}} \partial_x( \phi_1 + \phi_2) dx = 0, i=1,2,..,2n
\end{align}
where $\alpha_{2i-1}$ and $\alpha_{2i}$ are the mid points of region
$A_i$ and $\in B_i$, respectively, so that the integration region
includes the $i$-th dislocation (Fig. \ref{fig1sup}B). By definition
$\sum_iQ_i=0$ so that the number of independent constraints (for $n > 0$)
is $2n-1$. Each constraint reduces the number of states by a factor of $m+l$ since only the sector of $\phi_1+\phi_2$ is involved. Therefore the topological degeneracy in the presence of $n > 0$ pairs of dislocations is
\begin{align}
N=\frac{N_0}{(m+l)^{2n-1}}=\left|(m^2 - l^2) (m-l)^{n-1}\right|.
\end{align}

\section{More generic topological nematic states and effective topological field theory}
\label{generalTFT}

\begin{figure}
\centerline{
\includegraphics[width=2.5in]{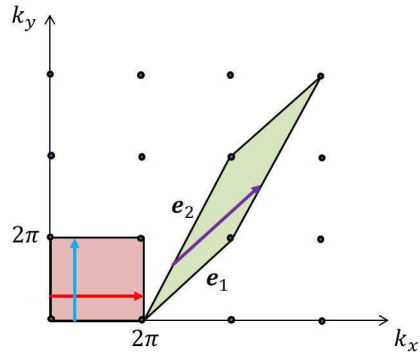}
}
\caption{
\label{fig:generalWan}
Illustration of different definitions of Wannier states, which lead to different types of topological nematic states. ${\bf e}_1$ and ${\bf e}_2$ are two reciprocal vectors defining a Brillouin zone. The Wannier state basis can be constructed by taking the Fourier transform of Bloch states along one periodic direction of the Brillouin zone, marked by the red, blue and purple lines with arrows. The red, blue and purple lines correspond to topological nematic states of the type $(1,0)$, $(0,1)$ and $(1,1)$, respectively (see text).}
\end{figure}

As shown above, the topological degeneracy is only associated with the $x$-dislocations
and not to $y$-dislocations. Such states apparently break rotation symmetry but preserve
translation symmetry, which is why we name them \it topological nematic states. \rm The reason
for the disparity is that the FQH states are constructed using Wannier states $\ket{W_{K}^{1,2}}$
localized in the $x$-direction. If we instead define the mapping from the lattice system to the bilayer FQH system
using Wannier states localized in the $y$-direction, the resulting state will have topological degeneracy associated
with $y$-dislocations. These two different types of topological states can be related by redefining the
Brillouin zone. More generally, we can choose the reciprocal lattice vectors to be ${\bf e}_1=(p,q)$ and
${\bf e}_2=(p',q')$ with $\left|{\bf e}_1\times {\bf e}_2\right|=pq'-qp'=1$, and define one-dimensional 
Wannier states by the Fourier transform of Bloch states along
the ${\bf e}_1$ direction, as shown in Fig. \ref{fig:generalWan}. The Wannier states obtained in this way are localized along ${\bf e}_1$ direction. For a dislocation with Burgers vector ${\bf b}=(b_x,b_y)$, around a dislocation the translation along the ${\bf e}_1$ direction is ${\bf b}\cdot {\bf e}_1=b_xp+b_yq$. Therefore the exchange of two layers and topological degeneracy only occurs when ${\bf b}\cdot {\bf e}_1=b_xp+b_yq=1$ mod $2$.
This condition classifies the topological nematic states through four types defined by $(p,q)~{\rm mod}~2=(0,0),(0,1),(1,0),(1,1)$.
The topological degeneracy is associated with $x$ ($y$)-dislocations if and only if $p$ ($q$) is odd. Rigorously speaking, $(0,0)$ cannot be realized by any state constructed in this way, because $pq'-qp'=1$ requires that at least one of $p$ and $q$ is odd. However one can interpret the type $(0,0)$ as the ordinary bilayer FQH state, where the two layers are not exchanged under any translation. 

To describe the topological nematic states better, we have developed
an effective field theory that naturally incorporates the interplay between topological properties and the lattice dislocations in this system.
Note that a dislocation in a two-dimensional crystal is a defect with long range interaction, similar to superfluid vortices.
Thus the effective field theory is different from the topological field theory studied in Ref. \cite{barkeshli2010}
which describes the phase with finite energy twist defects. The proper effective field theory should satisfy the following conditions:
\begin{enumerate}
\item The theory describes the dynamics of the crystal, which is characterized by the displacement field ${\bf u}=(u_x,u_y)$. In a 2d crystal the translation symmetry $R\times R$ is broken to $\mathbb{Z}\times \mathbb{Z}$, so that the order parameter ${\bf u}$ is periodic $u_x\sim u_x+1,~u_y\sim u_y+1$. Therefore ${\bf u}\in U(1)\times U(1)$.
    \item The theory should reduce to a
$U(1)\times U(1)$ Chern-Simons theory in the absence of dislocations.
\item  When dislocations are present, the two
$U(1)$ gauge groups should be exchanged when we take a reference point around a dislocation point.
\end{enumerate}

A natural effective theory satisfying the conditions above can be constructed by using a $U(2)$ gauge field
$a_\mu$ coupled to a Higgs field $H$ of the form $H  ={\bf \sigma}\cdot {\bf n}e^{i\theta({\bf u})}$,
with $\sigma$ are the three Pauli matrices, ${\bf n}$ a real unit vector, and
\begin{align}
\theta({\bf u})={\bf u}\cdot{\bf e}_1
\end{align}
is a phase factor determined by the displacement field ${\bf u}$ . Here ${\bf e}_1$ is the reciprocal vector defining the Wannier states as is illustrated in Fig. \ref{fig:generalWan} and earlier in this section.
$H$ is a traceless, unitary $2\times 2$ matrix in the adjoint representation of $U(2)$,
which transforms under a gauge transformation $H \rightarrow g^{-1} H g$, with $g$ a $U(2)$
matrix. $\bf{n}$ transforms as a vector under the $SU(2)$ subgroup of the $U(2)$ and $e^{i\theta}$ is gauge-invariant.
The effective Lagrangian has the following form: 
\begin{align}
\mathcal{L}=&\frac12\left[\rho\left(\partial_t {\bf u}\right)^2-\lambda_{ijkl}\partial_iu_j\partial_ku_l\right]\nonumber\\
&+\frac{m-l}{4\pi}\epsilon^{\mu\nu\tau}{\rm Tr}\left[a_\mu\partial_\nu a_\tau+\frac{2i}{3}a_\mu a_\nu a_\tau\right]
\nonumber \\
&+\frac{l}{4\pi}\epsilon^{\mu\nu\tau}{\rm Tr}\left[a_\mu\right]\partial_\nu {\rm Tr}\left[a_\tau\right]
+J{\rm Tr} \left[ D_\mu H^\dagger D_\mu H \right]
\label{Seff}
\end{align}
where the covariant derivative is $D_\mu H = \partial_\mu H + i\left[a_\mu, H \right]$. The first term is the standard elasticity theory of the crystal. The Chern-Simons fields describe the topological degrees of freedom of the electrons, which is coupled with the crystal through the Higgs field $H$.

We first consider a system without dislocation. For example for ${\bf u}=0$ one can choose a constant Higgs field
$H=\sigma_z$, which gives mass to two of the four components of the $U(2)$ gauge field $a$. The part of $a_\mu$ that remains massless is:
$a_\mu=a^0_\mu\frac{\bf 1}2+a^3_\mu\frac{\sigma^z}2$. Denoting
$a_\mu^{u(d)}=\frac12\left(a_\mu^0\pm a_\mu^3\right)$, the CS term for $a^u$ and $a^d$ reduces to
\begin{align}
\mathcal{L}_{CS}&=\frac{1}{4\pi}\epsilon^{\mu\nu\tau}\left(ma_\mu^u\partial_\nu a_\tau^u+ma_\mu^d\partial_\nu a_\tau^d+2la_\mu^u\partial_\nu a_\tau^d\right)
\end{align}
which correctly recovers the $U(1)\times U(1)$ CS theory of the $(mml)$ state \cite{wen1992kMatrix,wen04}, with $u$ and $d$ labeling the two layers.


Now we consider the effect of dislocations. Around a dislocation with Burgers vector ${\bf b}$, $\theta$ changes by $\pi{\bf e}_1\cdot{\bf b}$. Therefore if ${\bf e}_1\cdot {\bf b}=1$ mod $2$, $e^{i\theta}$ has a half winding around the dislocation. This is
allowed if and only if ${\bf n}$ also has a half-winding, compensating for the minus sign from the shift of $\theta$.
Such a topological defect is similar to the half vortex in a spinful $(p+ip)$ superconductor.\cite{chung2007} Different from the $p+ip$ superconductor in which the $SU(2)$ charge is global and the $U(1)$ charge is coupled to the electromagnetic gauge field, in the current case the $SU(2)$ charge carried by the vector ${\bf n}$ is gauged and the $U(1)$ charge remains global. Mathematically, the space of the order parameter described by $H$ is $S^2\times U(1)/Z_2$; the
$Z_2$ corresponds to identifying $e^{i\theta}{\bf n}$ with $e^{i(\theta+\pi)}(-{\bf n})$.
This has a nontrivial fundamental group: $\pi_1\left(S^2\times U(1)/Z_2\right)=Z_2$,
allowing topologically non-trivial point defects. The
invariant subgroup of $U(2)$ preserving the Higgs field is $U(1)\times U(1)$, defined by the rotations
$g=e^{i\left(\alpha+i\phi{\bf n\cdot\sigma}\right)/2}$. When we adiabatically take a point around a dislocation,
the vector ${\bf n}({\bf r})$ is adiabatically rotated to $-{\bf n}({\bf r})$, such that the two $U(1)$'s in the
invariant subgroup are exchanged. Thus the effective theory (\ref{Seff}) reproduces the
fact that the two $U(1)$ gauge fields, describing the two layers, in the $U(1)\times U(1)$ Chern-Simons theory are exchanged when a
reference point is taken around a dislocation. By construction, this theory also correctly describes the fact that only dislocations with ${\bf e}_1\cdot {\bf b}$ odd have topological degeneracy. Due to the gapless Goldstone modes described by displacement field ${\bf u}$, the dislocations have logarithmic interaction.


\section{Even/Odd Effect and Detection in Numerics}\label{sec:evenodd}

Even in the absence of dislocations, the topological nematic states exhibit unique topological characteristics that can aid
in their detection in numerical simulations. To see this, below we will focus on the $(mml)$ topological nematic states of type $(1,0)$, 
defined on a lattice with periodic boundary conditions.

\begin{figure}
\centerline{
\includegraphics[width=3in]{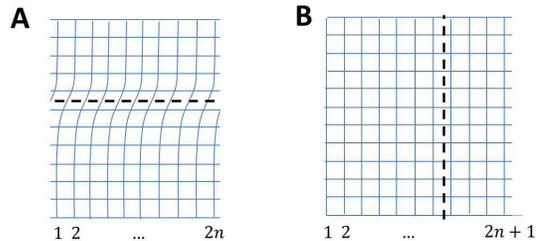}
}
\caption{
\label{fig:evenodd}
Illustration of two lattice configurations to detect the topological nematic states. (a) A torus with a kink in hopping along the dash line, and with even number of sites in $x$ direction. (b) A torus with no kink in hopping, but with odd number of sites in $x$ direction. In both cases, periodic boundary condition is imposed to both directions. For both configurations, the $(mml)$ topological nematic state of type $(1,0)$ has a reduced ground state degeneracy of $|m+l|$ instead of the degeneracy of $|m^2-l^2|$ on a torus with even number of sites in $x$ direction. }
\end{figure}

Let us first consider the situation shown in Fig. \ref{fig:evenodd} (a), where we suppose that along an entire line parallel to the $x$-direction,
the hoppings have been twisted. We also suppose that the number of sites in the $x$-direction is even.
Conceptually, we can imagine constructing this by taking a pair of dislocations,
moving one of them around the $x$-direction of the torus, and re-annihilating them. As we explain below, this change in the geometry effectively
changes the topology, as in the case with dislocations. To see this, consider the $(mm0)$ states, which can be viewed as
a direct product of two Laughlin $1/m$ states. For a regular lattice with periodic boundary conditions, the low energy theory
can be mapped onto two decoupled tori, each giving a degeneracy of $m$, for a total topological ground state degeneracy of
$m^2$. However, the twisted hopping has the effect of gluing the two tori together into a single torus, and so we expect that the
ground state degeneracy is actually $m$. We can further see this in an alternative way by using the edge theory picture as presented earlier.
By introducing twisted couplings into the edge theory, we introduce the tunneling terms
\begin{align}
\delta S_{twisted} = \sum_I \cos ( K_{IJ} \tilde{\phi}_J),
\end{align}
with $K = \left( \begin{matrix} m & l \\ l & m \\ \end{matrix} \right)$ and $\tilde{\phi}_J$ defined in Eq. (\ref{phitilde}). Naively, this would give $|\text{Det} K|$ different states labeled by the eigenstates of $e^{i(\tilde{\phi}_1\pm \tilde{\phi}_2)}=e^{ip_\pm/(m\pm l)}$, similar to the discussion in Sec. \ref{sec:edge}.
However, we observe that there is a gauge symmetry associated with the independent fermion number conservation within each layer, which
is associated with the shift symmetry
 \begin{align}
\phi_{Li} \rightarrow \phi_{Li} + \theta_i,
\nonumber \\
\phi_{Ri} \rightarrow \phi_{Ri} - \theta_i,
\end{align}
where $\theta_1$ and $\theta_2$ are independent. In the absence of tunneling between the different layers, this fermion number conservation
is clearly present. Since the $(mml)$ state is gapped, tunneling between the different layers is an irrelevant perturbation. Therefore
even in the presence of tunneling between the different layers, we may regard the fermion number conservation symmetry within each layer
as an emergent symmetry of the topological state. We conclude that, of the $|(m+l) (m-l)|$ degenerate minima of $\delta S_{twisted}$,
the ones associated with the same eigenvalue of $e^{i \tilde{\phi}_1 + i \tilde{\phi}_2}$ and different eigenvalues of $e^{i \tilde{\phi}_1 - i \tilde{\phi}_2}$ are gauge equivalent to each other. Therefore, we conclude that the topological degeneracy of such a
twisted lattice is $|m+l|$. In contrast, in the case of the non-twisted hoppings the gauge symmetry acts in a trivial
way on all the states and therefore does not affect the topological degeneracy $|{\rm Det}K|$.

Alternatively, consider a regular lattice, but with an odd number of sites along the $x$-direction, as shown in Fig. \ref{fig:evenodd} (b). In this case,
there will be a line at some point in $x$ where Wannier functions associated with different layers are adjacent. Cutting the system
along this vertical line, we obtain gapless edge states, and we observe that in gluing the system back together, we again have
twisted hoppings along the entire vertical line. Following the same argument as the last paragraph, in this case the ground state degeneracy is $|m+l|$.
This leads to a remarkable signature of the topological nematic states. For a regular lattice, when the number of sites in the $x$-direction is
even, the topological degeneracy is $|m^2 - l^2|$. When it is odd, the topological degeneracy is $|m+l|$. These topological
degeneracies are independent of the length in the $y$-direction. Such an even-odd
effect can easily be used as a numerical diagnostic to test for such topological nematic states.


\section{Domain Walls And Translation Symmetry-Protected Gapless Modes}

If the strongly interacting Hamiltonian that realizes the topological nematic states has a discrete lattice rotation symmetry,
then the topological nematic states associated with the class (1,0) or (0,1) (see Section \ref{generalTFT}) 
spontaneously break this discrete rotation symmetry. In such a situation, a physically realistic system
can be expected to consist of domains involving different orientations. At the domain walls, it is possible that translation symmetry
is preserved along the domain wall. As we show below, the translation symmetry along the domain wall can protect a single
bosonic channel from being gapped out.

\begin{figure}
\centerline{
\includegraphics[width=3in]{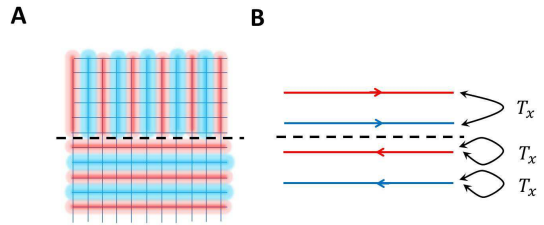}
}
\caption{
\label{fig:domainwall}
(a) Illustration of the domain wall between two topological nematic states $(1,0)$ (upper half) and $(0,1)$ (lower half). (b) Illustration that the left and right moving edge states along the domain wall transform differently in lattice translation $T_x$. The right movers from the $(1,0)$ state are exchanged by $T_x$ while the left movers from the $(0,1)$ state are separately translation invariant.}
\end{figure}

For concreteness, consider a domain wall between a $(0,1)$ and a $(1,0)$ topological nematic state, as shown in Fig. \ref{fig:domainwall}. Before coupling the two domains together,
each one has two branches of chiral gapless edge states: $\phi_{Li}$ and $\phi_{Ri}$, for $i = 1,2$, where $L/R$ stand for the
left/right-movers. Naively, these counterpropagating modes will be gapped out by inter-edge tunneling. However, they
have different properties under translation symmetry. In particular, under a translation along the domain wall,
\begin{align}
\phi_{L1} \rightarrow \phi_{L1} , \;\;\;& \phi_{L2} \rightarrow \phi_{L2},
\nonumber \\
\phi_{R1} \rightarrow \phi_{R2} , \;\;\;& \phi_{R2} \rightarrow \phi_{R1},
\end{align}
We see that for the translation symmetry along the domain
wall to be preserved, the only allowed inter-edge tunneling terms must involve the symmetric combination $\cos\left(\phi_{L1} + \phi_{L2}+\phi_{R1}+\phi_{R2}\right)$.
Therefore, either translation symmetry is preserved and the mode associated with the relative combination $\phi_{1L}+\phi_{1R} - \phi_{2L}-\phi_{2R}$
is gapless, or translation symmetry is spontaneously broken along the domain wall. This provides a novel example of
translation symmetry protected gapless edge states.

\section{Discussion and Conclusions}


This realization that lattice dislocations carry non-trivial topological degeneracy raises a host of possible
new directions. Conceptually, we are now in need of a more general theory of topological degeneracy of
dislocations and their possible braiding properties in general FQH states, both in bands
with higher Chern numbers and in non-Abelian FQH states. For example, for a Chern insulator with Chern number
$C_1=N$, the same construction maps the lattice system to a $N$-layer FQH state. Dislocations are "branch cuts" with
degree $N$ in such a system, around which the $N$ layers are cyclically permuted. It would also be important to verify our
predictions through numerical studies. The even-odd effect discussed in Sec. (\ref{sec:evenodd}) is expected to generalize to a $\text{mod } N$ behavior in the topological degeneracy for $C_1=N$.

Note that while the dislocations studied here have non-Abelian properties, they are
confined excitations of the system because the energy cost of separating dislocations grows logarithmically
with their distance. Nevertheless, physical materials can easily have many lattice dislocations, and if
FQH states are realized in a material with higher Chern bands, we expect that the lattice dislocations, in the limit
of low dislocation density, can induce an extensive topologically stable entropy for
temperatures $T \gtrsim \Delta e^{- l/\xi}$, where $\xi \propto 1/\Delta$ is the correlation length of the gapped FQH state,
$\Delta$ is the energy gap, and $l$ is the typical spacing between dislocations.

A further interesting, exotic possibility is to imagine quantum melting the lattice to deconfine these
dislocations. The resulting states, if realizable,
are ``topological liquid crystals" 
and may correspond to the non-Abelian orbifold states proposed in \cite{BW1121} and that are described by
$U(1) \times U(1) \rtimes Z_2$ CS theory.\cite{barkeshli2010}

\section*{Acknowledgements}

We would like to acknowledge inspiring discussions with Alexei Kitaev and Xiao-Gang Wen and thank the Simons Center for Geometry and Physics, Stony Brook University for hospitality where these discussions were made. We would like to acknowledge helpful discussions with Zheng-Cheng Gu, Steven A. Kivelson and Cenke Xu. This work was supported by a Simons Fellowship (MB) and the Alfred P. Sloan foundation and the David and Lucile Packard Foundation (XLQ). We thank the KITP program Topological Insulators and Superconductors for hospitality while part of this work was completed. This research was supported in part by the National Science Foundation under Grant No. NSF PHY11-25915.


\end{document}